\documentclass[12pt, draftclsnofoot, onecolumn]{IEEEtran}

\usepackage{xspace,amsmath,amssymb,amsfonts,epsfig,syntonly}
\usepackage{cite,bm,color,url,textcomp}
\usepackage{algorithmicx,algorithm,algpseudocode}
\usepackage{epstopdf}
\usepackage{empheq,multirow}
\usepackage{diagbox}
\usepackage{makecell}
\usepackage{array}

\newtheorem{lemma}{\hspace{-11pt}\bf Lemma}

\newtheorem{proposition}{\hspace{-11pt}\bf Proposition}

\long\def\symbolfootnote[#1]#2{\begingroup
\def\thefootnote{\fnsymbol{footnote}}
\footnote[#1]{#2}\endgroup}

\psfull

\allowdisplaybreaks[4]

\begin{document}
\title{Cooling-Aware Resource Allocation and Load Management for Mobile Edge Computing Systems}

\author{Xiaojing Chen,~\IEEEmembership{Member, IEEE},  Zhouyu Lu, Wei Ni,~\IEEEmembership{Senior Member, IEEE},\\ Xin Wang,~\IEEEmembership{Senior Member, IEEE}, Feng Wang,~\IEEEmembership{Member, IEEE}, \\Shunqing Zhang,~\IEEEmembership{Senior Member, IEEE}, and Shugong Xu,~\IEEEmembership{Fellow, IEEE}
\thanks{Work in this paper was supported by the National Natural Science Foundation of China grants 61701293, 61871262, 61901251 and 61671154, the National Science and Technology Major Project grant 2018ZX03001009, and the National Key R\&D Program of China grant 2017YFE0121400.

X. Chen, Z. Lu, S. Zhang and S. Xu are with the Shanghai Institute for Advanced Communication and Data Science, Shanghai University, Shanghai 200444, China. Emails:~\{jodiechen, zhouyulu, shunqing, shugong\}@shu.edu.cn.

W. Ni is with the Commonwealth Scientific and Industrial Research Organization (CSIRO), Sydney, NSW 2122, Australia. Email: wei.ni@data61.csiro.au.

X. Wang is with the State Key Laboratory of ASIC and System, the Shanghai Institute for Advanced Communication and Data Science, the Department of Communication Science and Engineering, Fudan University, Shanghai 200433, China. Email: xwang11@fudan.edu.cn.

F. Wang is with the School of Information Engineering, Guangdong University of Technology, Guangzhou 510006, China. Email:
fengwang13@gdut.edu.cn.

}}
%

\maketitle


\setcounter{page}{1}
\begin{abstract}
Driven by explosive computation demands of Internet of Things (IoT), mobile edge computing (MEC) provides a promising technique to enhance the computation capability for mobile users.
In this paper, we propose a joint resource allocation and load management mechanism in an MEC system with wireless power transfer (WPT), by jointly optimizing the transmit power for WPT, the local/edge computing load, the offloading time, and the frequencies of the central processing units (CPUs) at the access point (AP) and the users.
To achieve an energy-efficient and sustainable WPT-MEC system, we minimize the total energy consumption of the AP, while meeting computation latency requirements. Cooling energy which is non-negligible, is taken into account in minimizing the energy consumption of the MEC system. By rigorously orchestrating the state-of-the-art optimization techniques, we design an iterative algorithm and obtain the optimal solution in a semi-closed form. Based on the solution, interesting properties and insights are summarized. Extensive numerical tests show that the proposed algorithm can save up to 90.4\% the energy of existing benchmarks. 
\end{abstract}


\section{Introduction}
\label{sec:intro}
The recent development of the Internet of Things (IoT) has ushered in many new mobile applications \cite{chen2019learning,zhang2019first,zhang17fundamental,survey2018parvez}. New applications, such as autonomous driving, interactive online gaming, and augmented reality (AR), require low-latency communication and intensive computation for massive mobile devices \cite{joint19feng}. Critical challenges arise from providing quality of experience (QoE) for these applications in conventional communication systems \cite{chen2015provisioning, Chen2016Optimal}, where mobile devices are generally equipped with limited computation capability.
The emerging mobile edge computing (MEC) offers a new paradigm to offload computing workload from mobile users to network edge \cite{Mao2017A,mobile2018zhang}. In MEC systems, edge facilities, e.g., base stations (BSs) and access points (APs), are integrated with MEC servers. This enhances computation capability for mobile users by leveraging remote execution, and significantly reduces communication latency by having MEC servers placed in the proximity of mobile users \cite{mobile2018abbas}.
By integrating the network function virtualization (NFV) techniques, MEC is expected to provide flexibility on resource scheduling and service deployment~\cite{chen18m, chen19auto}.

Achieving energy-efficient and self-sustainable computing for massive computation-intensive devices is challenging to the IoT. Energy harvesting (EH) powered MEC designs have been increasingly studied in the literature.
In general, there are three typical types of EH mechanisms in MEC systems: APs are powered by renewable energy  \cite{xu2016online}, \cite{guo2018joint},  mobile users are powered by renewable energy \cite{zhang2018energy,zhang2018distributed,mao2016dynamic,min2019learning,wei2018dynamic}, or wireless power transfer (WPT) is performed from AP to IoT devices via radio-frequency (RF) signals \cite{wang18dynamic, optimal19feng, joint18wang, zhang18EE}.
In \cite{xu2016online}, a system operator learns online the central processing unit (CPU) cycles of the MEC server and the amount of tasks to be offloaded from the server to the central cloud, adapting to the renewable energy and the congestion states of the core network.
Considering inter-cell interference in dense small cells, a distributed three-stage strategy was proposed in \cite{guo2018joint} to jointly optimize task offloading, channel allocation, and computing resource allocation.
Focusing on practical random system environments, algorithms based on Lyapunov optimization techniques \cite{zhang2018energy,zhang2018distributed,mao2016dynamic} and deep learning \cite{min2019learning,wei2018dynamic} were developed to yield dynamic load distribution and resource allocation strategies.

In addition, RF signal based WPT is a promising technique to balance energy distribution and provide sustainable energy supply for wireless devices \cite{xu2014energy}.
WPT is able to provide a flexible and stable EH method, as it can be performed regardless of time, place and weather condition \cite{hu2020modeling}.
Dynamic interface-selection and resource allocation \cite{wang18dynamic}, and joint optimization of energy transmit beamforming, computation and wireless resource allocation \cite{optimal19feng,joint18wang} have been investigated for WPT-MEC systems.
In \cite{zhang18EE}, the energy cost of a multiuser MEC system with simultaneous wireless information and power transfer (SWIPT) was minimized while meeting the latency requirement.


The existing works on EH powered MEC systems have overlooked the cooling energy of the AP, which accounts for a non-negligible proportion in the system energy consumption. Recently, the models of cooling energy have been developed to facilitate the energy-efficient design of data centers \cite{chen2015cooling,robust16tianyi,chen20joint}. Two cooling modes have been typically considered: outside-air (OA) and chilled-water (CW) cooling. As the capability of OA cooling relies on and is limited by the air temperature, CW cooling generally serves as a complementary approach for OA cooling.

In this paper, by further considering the AP's cooling energy consumption, we jointly optimize resource allocation and load management for cooling-aware WPT-MEC systems, where a multi-antenna AP with stable power supply coordinates the WPT to multiple users. The users first harvest energy by WPT from the AP, and then rely on the energy to execute their tasks via local computing and computation offloading. Suppose that the users' tasks are independent and partitionable, i.e., each task can be partitioned into two parts for local computing and offloading, respectively \cite{mobile2017mach}.
The main contributions of this paper are summarized as follows.
\begin{itemize}
\item We minimize the total energy consumption of the AP to achieve an energy-efficient and sustainable WPT-MEC system, while satisfying computation latency constraints.  
    For the first time, the non-negligible cooling energy is involved in the energy consumption model. Coupling energy model between computing and cooling management is presented. 
    We design an iterative algorithm to jointly manage edge computing and cooling activities, and obtain the optimal solution in a semi-closed form.

\item  By orchestrating the alternating optimization technique and Lagrange duality method, we jointly optimize the transmit power for WPT, the local/edge computing load, the offloading time, and the CPU frequencies at the AP and at the users. A joint design of resource allocation and load management is developed to achieve a balance between the WPT supply from the AP and the energy demand for computation and offloading at the end users.
    Interesting properties and engineering insights are summarized. It is revealed that less offloading time is allowed with a stringent latency requirement.

\item Extensive numerical tests corroborate that the proposed joint design can substantially save the energy consumption of the AP, as compared to benchmark schemes. The proposed algorithm can save up to 90.4\% and 81.0\% of the energy, as compared to the full and half offloading benchmark schemes, respectively. It is revealed that the users would offload more data when the computation latency constraint becomes more stringent. It is also shown that the transmit power for WPT is affected by the channel power gain, the energy conversion efficiency and the scheduling time.
\end{itemize}

The rest of this paper is organized as follows. In Section II, the system model is introduced. In Section III, the proposed joint resource allocation and load management scheme is developed. We summarize the properties and insights of the proposed design in Section IV. Numerical tests are provided in Section V, followed by conclusions in Section VI. Notations in the paper are listed in Table I.

\renewcommand\arraystretch{0.8}
\begin{table*}[htbp]
	\caption{Notation and Definition}
	\label{tab:parameters}
	\begin{tabular}{p{2cm}|p{12cm}}
		\hline
		\emph{Notation} & \emph{Definition}\\
		\hline\hline
		$I,i,{\cal I}$ & Number, index and set of single-antenna mobile users\\
		\hline
		$T$ & Duration of the scheduling time slot\\
		\hline
		$\phi$  & Time splitting ratio for WPT\\
		\hline
		$R_i$ & Task size (in nats) arrived at user $i$\\
		\hline
		$a_i$ & Task ratio for local computing of user $i$\\
		\hline
		$B_i$ & Required number of CPU cycles for per nat computing by user $i$\\
		\hline
		$f_{u,i}$ & CPU frequency of user $i$\\
		\hline
		$f_{u,i}^{\max}$ & Maximum achievable CPU frequency of user $i$\\
		\hline
		$f_{s,i}$ & CPU frequency of the MEC server allocated to computing user $i$'s task\\
		\hline
		$f_{s}^{\max}$ & Maximum achievable CPU frequency of the server\\
		\hline
		$k_i$,  $\delta$ & Effective switched capacitance coefficient depending on the chip architecture at user $i$ and at the server\\
		\hline
		$\sigma^2$ & Noise power at the receiver of the AP\\
		\hline
		$g_i$ & Effective channel gain from user $i$ to the AP\\
		\hline	
		$H_i$ & Channel gain from the AP to user $i$\\
		\hline	
		$W$ & System bandwidth\\
		\hline
		$w_i$ & Uplink bandwidth allocated to user $i$\\
		\hline	
		$\theta_i$ & User $i$'s energy conversion efficiency\\
		\hline
		$E_{\text{loc},i}$ & Energy consumption for local computing at user $i$\\
		\hline
		$P_{\text{loc},i}$ & Power consumption for local computing at user $i$\\
		\hline
		$T_{\text{loc},i}$ & Task execution time for local computing at user $i$\\
		\hline
		$E_{\text{off},i}$ & Energy consumption for computation offloading at user $i$\\
		\hline
		$P_{\text{tra},i}$ & Transmit power for computation offloading of user $i$\\
		\hline
		$T_{\text{off},i}$ & Offloading time of user $i$\\
		\hline
		$E_{\text{comp},i}$ &  Energy consumption for executing user $i$'s task at the AP\\
		\hline
		$P_{\text{comp},i}$ & Power consumption for executing user $i$'s task at the server\\
		\hline
		$T_{\text{exe},i}$ &  Execution time for computing user $i$'s task at the AP\\
		\hline
		$P_{\text{OA}}$ &  Power consumption of OA cooling\\
		\hline
		$P_{\text{CW}}$ & Power consumption of CW cooling\\
		\hline
		$P_{\text{cool}}$ & Cooling power consumption\\
		\hline
		$E_{\text{cool}}$ & Cooling energy consumption at the AP\\
		\hline
		$E_{h,i}$ & Accumulative energy harvested by user $i$ from the AP's RF signals at the end of Phase-I\\
		\hline
		$P_{b,i}$ & Transmit power for WPT allocated to user $i$\\
		\hline
		$P_b^{\max}$ & Maximum transmit power of the AP\\
		\hline	
		$\lambda_i$, $\mu_i$, $\pi $, $\nu$ & Lagrange multipliers\\
		\hline
		$\boldsymbol{\omega}$ & $\boldsymbol{\omega}:=\{\lambda_i, \mu_i,\pi ,\nu,\forall i \in {\cal I}\}$ \\
		\hline
		$n$ & Iteration index\\
		\hline
		$g_{\lambda_i}(n)$, $g_{\mu_i}(n)$, $g_{\pi }(n)$, $g_{\nu}(n)$ & Subgradients\\
		\hline
		$\boldsymbol g$ & ${\boldsymbol g}:=\{g_{\lambda_i}(n),g_{\mu_i}(n), g_{\pi }(n), g_{\nu}(n)\}$ \\
		\hline
		$\eta$ & Appropriate stepsize\\
		\hline
	\end{tabular}
\end{table*}

\section{System Models}

\begin{figure}
	\centering
	\includegraphics[width=0.7\textwidth]{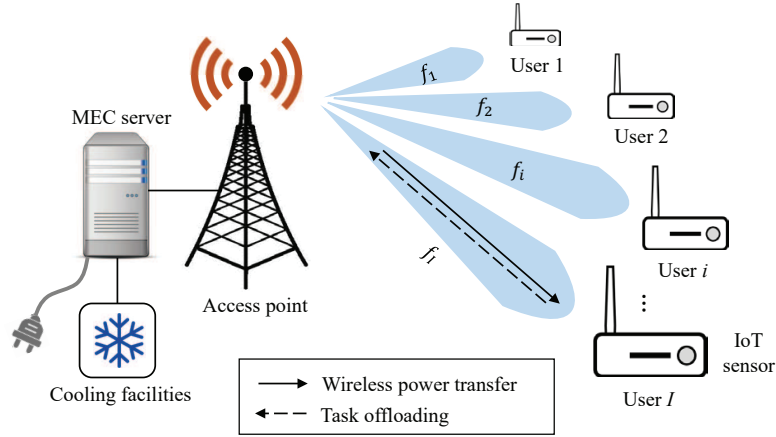}
	\caption{A WPT-MEC system with cooling facilities.}
	\label{structure1}
\end{figure}

\begin{figure}
\centering
\includegraphics[width=0.6\textwidth]{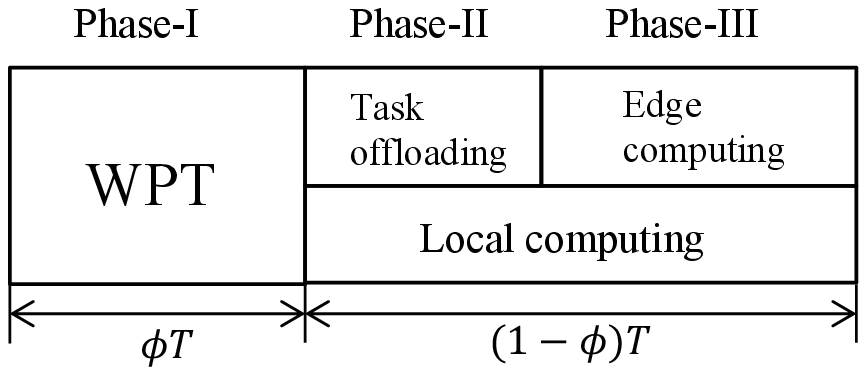}
\caption{The cooperation between the AP and the end users in the WPT-MEC system.}
\label{model1}
\end{figure}
Consider a wireless-powered multiuser MEC system, where a multiple-antenna AP with persistent power supply communicates with a set of single-antenna mobile users ${\cal I}:=\{1,\ldots, I\}$ in the presence of dominating line-of-sight (LoS) between the AP and the users; see Fig.~\ref{structure1}. The AP can perform coherent beamforming for WPT to users \cite{Ko2008Coherent}. The AP is integrated with an MEC server for task computing. The users are powered by rechargeable batteries which need to be charged by the RF signals of the AP. The physical framework between the AP and the users in the WPT-MEC system is illustrated in Fig.~\ref{model1}. In a slot with a duration of $T$, each user first scavenges energy from the AP in Phase I, and then utilizes the energy to perform local computing and partial computation offloading (if needed). The uplink computation offloading is performed in Phase II, followed by task execution at the AP in Phase III.
Suppose that computation offloadings by different users are performed simultaneously over orthogonal frequency channels. For simplicity, the system bandwidth $W$ is assumed to be equally divided among the users, i.e., we have $w_i=W/I$, where $w_i$ is the uplink bandwidth allocated to user $i$.
The users can also execute the remaining tasks locally with the harvested energy throughout Phases II and III. Let $\phi T$ denote the duration of Phase I, where $ \phi \in [0, 1]$ is the time splitting ratio for WPT. Also, let $T_{\text{off},i}$ denote the offloading time of user $i$ (i.e., the duration of Phase II). Then, the duration of Phase III is $(1-\phi)T-T_{\text{off},i}$.
After edge computing, the AP returns the computation results to the respective users. Since the results are usually of small size, we ignore the result downloading time, as well as the energy consumption for transmitting and receiving the computation results \cite{joint18wang}.
The slot length $T$ is set to be no larger than the latency requirements of MEC tasks from different users and the wireless channel coherence time, so that the channels remain time-invariant during a slot. Assume that the task amount to be executed per user and the channel state information (CSI) from the $I$ users are known at the AP\footnote{CSI estimation techniques for WPT systems have been extensively studied in the literature, e.g. \cite{zeng2014optimized} and \cite{xu2016general}.  CSI estimation is out of scope of this paper.}. The AP coordinates the downlink WPT, the computation offloading, and the edge and local computing for the WPT-MEC system.

\subsection{Task Execution Time and Energy Consumption}
\subsubsection{Energy transfer and harvesting}
Let $E_{h,i}$ denote the total energy harvested by user $i$ from the RF signals of the AP by the end of Phase I. $E_{h,j}$ is given by:
\begin{equation}
E_{h,i}=P_{b,i} \phi T \theta_i H_i,
\end{equation}
where $P_{b,i}$ is the transmit power of the AP assigned to user~$i$, e.g., by coherent beamforming. $\sum_i P_{b,i}\leq P_b^{\max}$ with $P_b^{\max}$ being the maximum transmit power of the AP.
$\theta_i\in (0,1)$ is user $i$'s energy conversion efficiency. $H_i$ is the channel gain from the AP to user $i$.

\subsubsection{Local computing}
Let $R_i$ denote the task size (in nats) at user $i$ and to be processed by $T$, and $a_i \in [0,1]$ denote the ratio of the task to be locally executed.
Let $B_i$ denote the required number of CPU cycles for per-nat computing by user $i$. Then the total number of CPU cycles required for accomplishing user $i$'s task is $a_i R_i B_i$. By leveraging dynamic voltage and frequency scaling (DVFS) techniques \cite{Mao2017A}, user $i$ can adjust the energy consumption for local computing by configuring a variable CPU frequency $f_{u,i} \in [0, f_{u,i}^{\max}]$, where $f_{u,i}^{\max}$ is the maximum achievable CPU frequency of user $i$.
The task execution time for local computing at user $i$ is then given by:
\begin{equation}\label{Tloc}
T_{\text{loc},i}=  \frac{a_i R_i B_i}{f_{u,i}},\quad \forall i \in {\cal I}.
\end{equation}
Accordingly, the energy consumption for local computing at user $i$ is:
\begin{equation}
E_{\text{loc},i}= P_{\text{loc},i}T_{\text{loc},i}=a_i R_i B_i k_i f_{u,i}^2,\quad \forall i \in {\cal I},
\end{equation}
where $P_{\text{loc},i}= k_i f_{u,i}^3$ is the power consumption for local computing at user $i$, and $k_i$ is the effective switched capacitance coefficient depending on the chip architecture at user $i$ \cite{burd1996processor}.

\subsubsection{Task offloading}
After the energy-harvesting phase (i.e., Phase I), user $i$ can offload $(1-a_i)R_i$ of its tasks to the AP for edge computing.
The energy consumption for computation offloading at user $i$ is:
\begin{equation}
E_{\text{off},i}=  P_{\text{tra},i} T_{\text{off},i}, \quad \forall i \in {\cal I},
\end{equation}
where $P_{\text{tra},i}$ is the transmit power of user $i$ for computation offloading, as given by
\begin{equation}\label{power_transmit}
P_{\text{tra},i} =\frac{\sigma^2}{g_i}(e^{\frac{(1-a_i)R_i}{T_{\text{off},i}w_i}}-1),
\end{equation}
where $\sigma^2$ is the noise power at the AP, and $g_i$ is the effective channel gain from user $i$ to the AP. 

\subsubsection{Edge computing}
In Phase III, the AP performs edge computing for the tasks received from its users. Let $f_{s,i}$ denote the CPU frequency of the MEC server allocated to computing user $i$'s task. We have $\sum_i f_{s,i} \leq f_{s}^{\max}$ with $f_{s}^{\max}$ being the CPU capacity of the server.
The execution time for computing user $i$'s task at the AP can be given by:
\begin{equation}\label{time_comp}
T_{\text{exe},i}= \frac{(1-a_i)R_iB_i}{f_{s,i}}, \quad \forall i \in {\cal I}.
\end{equation}
The corresponding energy consumption of the AP is:
\begin{equation}\label{power_comp}
E_{\text{comp},i} =  P_{\text{comp},i} T_{\text{exe},i} =(1-a_i)R_i B_i \delta f_{s,i}^2, \quad \forall i \in {\cal I},
\end{equation}
where $P_{\text{comp},i}=\delta f_{s,i}^3$ is the power consumption for executing user $i$'s task at the server, and $\delta$ is the effective switched capacitance coefficient depending on the chip architecture at the MEC server \cite{burd1996processor}.


\subsubsection{Cooling energy demand}
With the increase of computation demand at the MEC server, a non-negligible amount of energy is consumed to cool it off. There are generally two cooling modes \cite{robust16tianyi}: OA cooling and CW cooling.
For OA cooling, energy is predominantly consumed by blowers, and can be modeled as a cubic function of the blower speed \cite{Zhou2012Optimization}. The blower speed is linear to the total computing power of the server $\sum_i P_{\text{comp},i}$. Therefore, the OA cooling power can be modeled as a cubic function of $\sum_i P_{\text{comp},i}$ \cite{Zhou2012Optimization}; that is,
\begin{equation}\label{OA}
P_{\text{OA}}=\epsilon_1 \Big(\sum_i P_{\text{comp},i}\Big)^3,  \quad  \sum_i  P_{\text{comp},i} \leq P_{a}^{\max},
\end{equation}
where $\epsilon_1>0$ depends on the temperature difference $\Delta_t$ between the hot exhausting air from the server racks and the environment air temperature. $P_{a}^{\max}=C \Delta_t$ is the capacity of OA cooling, where $C>0$ depends on the maximum OA flow rate. As the capability of OA cooling depends on the air temperature, this approach is generally complemented by other more stable approaches, such as CW cooling.

Based on the actual measurement of a chiller, the power consumption of CW cooling is modeled to be linear to the computing power of the server $\sum_i P_{\text{comp},i}$ \cite{robust16tianyi}, as given by
\begin{equation}\label{CW}
P_{\text{CW}}=\epsilon_2 \sum_i P_{\text{comp},i},
\end{equation}
where $\epsilon_2>0$ is a constant depending on the chiller's characteristics.

With different power consumptions and capabilities of the two cooling mechanisms, we take both into consideration and aim to find the most energy-efficient combination of OA and CW cooling \cite{robust16tianyi}. Let $P_{a}$ and $P_{c}$ denote the amounts of the MEC server's power consumption distributed between OA and CW cooling, respectively (i.e., $P_{a}+P_{c}=\sum_i P_{\text{comp},i}$). According to \eqref{OA} and \eqref{CW}, given $\sum_i P_{\text{comp},i}$, the minimum power consumption for cooling is \cite{robust16tianyi}
\begin{align}\label{cooling}
P_{\text{cool}}&= \min_{P_{a}} ~~(P_{\text{OA}}+P_{\text{CW}}) \notag\\
&= \min_{P_{a}} ~~\Big(\epsilon_1P_{a}^3+\epsilon_2 \big[\sum_i P_{\text{comp},i}-P_{a}\big]^+\Big) .
\end{align}
We then solve \eqref{cooling} and obtain the minimum cooling power consumption, as given by
\begin{align}\label{cooling1}
P_{\text{cool}}=\left\{
\begin{array}{ll}
    \epsilon_1(\sum_i P_{\text{comp},i})^3,~~\text{if}~\sum_i P_{\text{comp}}\leq P_{\text{TH}},\\
     \epsilon_1P_{\text{TH}}^3+\epsilon_2 [\sum_i P_{\text{comp},i}-P_{\text{TH}}]^+, ~~\text{otherwise},\\
\end{array}
\right.
\end{align}
where $P_{\text{TH}} := \min \{P_{a}^{\max}, \sqrt{\epsilon_2 /3\epsilon_1}\}$ denotes a temperature-based threshold. From \eqref{cooling1}, we see that if the total computing power is lower than $P_{\text{TH}}$, only OA cooling is employed; otherwise, both cooling mechanisms are applied and CW cooling is served as a complementary way.

Therefore, the cooling energy consumption stemming from executing users' tasks at the AP is given by:
\begin{equation}\label{cooling2}
E_{\text{cool}}=P_{\text{cool}} (1-\phi)T.
\end{equation}

Based on the definition of the cooling energy consumption in \eqref{cooling2}, we establish the following lemma.
\begin{lemma}\label{lemma1}
 The cooling energy consumption $E_{\text{cool}}(f_{s,i})$ is convex in $f_{s,i}$.
\end{lemma}

\begin{IEEEproof}
Function $P_{\text{comp},i}(f_{s,i})=\delta f_{s,i}^3$ is convex in $f_{s,i}\geq 0$. Hence, $E_{\text{cool}}(P_{\text{comp},i}(f_{s,i}))$ is convex in $f_{s,i}$ since $E_{\text{cool}}(P_{\text{comp},i})$ is convex and nondecreasing \cite{convex}.
\end{IEEEproof}


%

\section{Joint Resource Allocation and Load Management}
\subsection{Problem Formulation}
In this section, we design a joint resource allocation and load management mechanism for the cooling-aware WPT-MEC system. Consider the system's total energy minimization problem subject to energy causality and computation latency constraints. Specifically, we aim to minimize the energy consumption (for WPT, edge computing, and cooling) of the AP, while guaranteeing that: i) the energy consumed (by task offloading and local computing) at each user does not exceed the energy harvested, and ii) all the arrived tasks are successfully executed within a slot of duration $T$. To achieve this, we make joint decisions on the optimal transmit power for WPT ${\boldsymbol P_b}:=\{P_{b,1}, \ldots, P_{b,I}\}$, the local computing ratio ${\boldsymbol a}:=\{a_1, \ldots, a_I\}$, the CPU frequencies at the AP ${\boldsymbol f_s}:=\{f_{s,1}, \ldots, f_{s,I}\}$ and at the users ${\boldsymbol f_u}:=\{f_{u,1}, \ldots, f_{u,I}\}$, and the offloading time ${\boldsymbol T_{\text{off}}}:=\{T_{\text{off},1}, \ldots, T_{\text{off},I}\}$ for each user. The energy minimization problem of interest is formulated as follows:
\begin{subequations}\label{eq.prob}
\begin{align}
& \mathop{\text{min}}\limits_{\{{\boldsymbol a},{\boldsymbol f_s},{\boldsymbol f_u},{\boldsymbol P_b},{\boldsymbol T_{\text{off}}}\}} \; \sum_i \big(\frac{E_{h,i}}{\theta_iH_i}+E_{\text{comp},i}\big)+E_{\text{cool}} \label{eq.proba} \\
& \text{subject to:} \notag \\
& E_{\text{loc},i}+E_{\text{off},i}\leq E_{h,i}, \label{eq.probb}\\
  & T_{\text{off},i}+T_{\text{exe},i}\leq (1-\phi)T, \label{eq.probd}\\
  &T_{\text{loc},i}\leq (1-\phi)T,\label{eq.probe}\\
  &f_{u,i}\leq f_{u,i}^{\max}, \label{eq.probf}\\
  &\sum_i f_{s,i} \leq f_s^{\max}, \label{eq.probg}\\
  &\sum_i P_{b,i}\leq P_b^{\max}, ~~\forall i \in {\cal I}. \label{eq.probh}
\end{align}
\end{subequations}
The objective \eqref{eq.proba} is the total energy consumption of the AP. Constraint \eqref{eq.probb} is the energy causality constraint which ensures that the energy consumed at each device cannot exceed the energy harvested by the device through WPT. \eqref{eq.probd}-\eqref{eq.probe} are latency constraints, which guarantee that the tasks offloaded to the server for edge computing, and those remaining for local computing must be executed by the end of a slot. \eqref{eq.probf}-\eqref{eq.probh} are self-explanatory.

\subsection{Proposed Joint Resource and Load Management Design}
Problem \eqref{eq.prob} is non-convex due to the non-convex nature of \eqref{eq.proba}-\eqref{eq.probd}. Yet, from \eqref{power_comp}, we observe that the computing energy consumption, $E_{\text{comp},i}(a_i,f_{s,i})$, is convex in $f_{s,i}$ given $a_i$, and linear to $a_i$ given $f_{s,i}$ \cite{convex}. Therefore, we can resort to the alternating optimization technique to solve \eqref{eq.prob}. Specifically, we first calculate $\{{\boldsymbol f_s},{\boldsymbol f_u},{\boldsymbol P_b},{\boldsymbol T_{\text{off}}}\}$ given ${\boldsymbol a}$; and then determine ${\boldsymbol a}$ given $\{{\boldsymbol f_s},{\boldsymbol f_u},{\boldsymbol P_b},{\boldsymbol T_{\text{off}}}\}$. This process repeats until the convergence to a locally optimal solution to \eqref{eq.prob}. 

The following lemma is put forth to reveal the property of the local CPU frequency at the users ${\boldsymbol f_u}:=\{f_{u,1}, \ldots, f_{u,I}\}$ in the solution to \eqref{eq.prob}.
\begin{lemma}\label{lemma2}
The optimal solution for the local CPU frequency at the users ${\boldsymbol f_u}:=\{f_{u,1}, \ldots, f_{u,I}\}$ to problem (\ref{eq.prob}) satisfies
\begin{equation}
f_{u,i}=\frac{a_iR_i B_i}{(1-\phi)T}\leq f_{u,i}^{\max}.
\label{f_u}
\end{equation}
\end{lemma}
\begin{IEEEproof}
See Appendix A.
\end{IEEEproof}

Given the value of ${\boldsymbol a}$, problem \eqref{eq.prob} can be reformulated as the following problem \eqref{eq.prob1} by plugging \eqref{f_u}, i.e., $f_{u,i}$, to \eqref{eq.prob}, according to Lemma~\ref{lemma2}.
\begin{subequations}\label{eq.prob1}
\begin{align}
& \mathop{\text{min}}\limits_{\{{\boldsymbol f_s},{\boldsymbol P_b},{\boldsymbol T_{\text{off}}}\}} \; \sum_i \big(\phi T P_{b,i}+(1-a_i)R_i B_i \delta f_{s,i}^2\big)+E_{\text{cool}}  \label{eq.proba1} \\
& \text{subject to:} \notag \\
& \frac{a_i^3 R_i^3k_iB_i^3}{(1-\phi)^2 T^2}+\frac{\sigma^2}{g_i}(e^{\frac{(1-a_i)R_i}{T_{\text{off},i}w_i}}-1)T_{\text{off},i}\leq P_{b,i} \phi T \theta_i H_i, \label{eq.probb1}\\
  & T_{\text{off},i}+\frac{(1-a_i)R_iB_i}{f_{s,i}} \leq (1-\phi)T, \label{eq.probd1}\\
  & \sum_i f_{s,i}\leq f_s^{\max},\label{eq.probf1}\\
  &\sum_i P_{b,i}\leq P_b^{\max}, ~~\forall i \in {\cal I}. \label{eq.probh1}
\end{align}
\end{subequations}

Based on Lemma~\ref{lemma1}, the objective function \eqref{eq.proba1} is convex in $\{{\boldsymbol f_s},{\boldsymbol P_b}\}$. Let
\begin{equation}\label{eq.perspective}
P(\frac{(1-a_i)R_i}{T_{\text{off},i}}):=e^\frac{(1-a_i)R_i}{T_{\text{off},i}w_i}-1,
\end{equation}
which is a convex function of $\frac{(1-a_i)R_i}{T_{\text{off},i}}$, and its perspective function $P(\frac{(1-a_i)R_i}{T_{\text{off},i}})T_{\text{off},i}$ is also convex in $T_{\text{off},i}$ \cite{convex}. Since the objective function and all the constraints are convex in (\ref{eq.prob1}), problem (\ref{eq.prob1}) is convex and can be solved by general-purpose convex optimization solvers (e.g., the interior-point method~\cite{convex}).

In this paper, we propose to employ the Lagrange duality method to solve the problem, and obtain a semi-closed-form solution to provide engineering insights for joint resource allocation and load management in WPT-MEC systems.
Let $\{\lambda_i\}$, $\{\mu_i\}$, $\{\nu\}$ and $\{\pi\}$ denote the Lagrange multipliers associated with constraints \eqref{eq.probb1}, \eqref{eq.probd1}, \eqref{eq.probf1} and \eqref{eq.probh1}, respectively. By using $\boldsymbol{\omega}:=\{\lambda_i, \mu_i,\nu,\pi,\forall i \in {\cal I}\}$ to collect all the Lagrange multipliers, the Lagrangian function of \eqref{eq.prob1} is given by
\begin{align}
&{\cal L}({\boldsymbol f_s},{\boldsymbol P_b},{\boldsymbol T_{\text{off}}},\boldsymbol{\omega})\notag\\
&=\sum_i[\phi T P_{b,i} + (1-a_i)R_i\delta B_i f_{s,i}^2 ]+ E_{\text{cool}}(f_{s,i})\notag\\
&+\sum_i \lambda_i[\frac{a_i^3 R_i^3k_iB_i^3}{(1-\phi)^2 T^2}+\frac{\sigma^2}{g_i}(e^{\frac{(1-a_i)R_i}{T_{\text{off},i}w_i}}-1)T_{\text{off},i}- P_{b,i} \phi T \theta_i H_i] \notag\\
&+\sum_i \mu_i[T_{\text{off},i}+\frac{(1-a_i)R_iB_i}{f_{s,i}}- (1-\phi)T]\notag\\
&+\nu[\sum_i f_{s,i}- f_s^{\max}]+\pi[\sum_i P_{b,i}- P_b^{\max}].
\end{align}
The Lagrange dual function is given by
\begin{align}\label{eq.probdualfunc}
{\cal D}(\boldsymbol{\omega}):= \min\;\; {\cal L}({\boldsymbol f_s},{\boldsymbol P_b},{\boldsymbol T_{\text{off}}},\boldsymbol{\omega})
\end{align}
As a result, the dual problem of \eqref{eq.prob1} is given by
\begin{align}\label{eq.probdual}
\max \;\;\;&{\cal D}(\boldsymbol{\omega}) \notag\\
 \text{s.t.} \quad &\lambda_i\geq 0, \mu_i \geq 0, \quad\forall i \in {\cal I},\notag \\
&\nu \geq 0,  \pi \geq 0.
\end{align}


For the dual problem \eqref{eq.probdual}, a standard subgradient method can be employed to obtain the optimal $\boldsymbol{\omega}$. This amounts to running the following iterations:
\begin{subequations}\label{eq.multiplier}
\begin{align}
\lambda_i(n+1) &= [\lambda_i(n) + \eta g_{\lambda_i}(n)]^+, \quad\forall i \in {\cal I}, \\
\mu_i(n+1) &= [\mu_i(n) + \eta g_{\mu_i}(n)]^+, \quad\forall i \in {\cal I}, \\
\nu(n+1) &= [\nu(n) + \eta g_{\nu}(n)]^+, \\
\pi(n+1) &= [\pi(n) + \eta g_{\pi}(n)]^+ ,
\end{align}
\end{subequations}
where $n$ is the iteration index, $\eta >0$ is an appropriate stepsize, and ${\boldsymbol g}:=\{g_{\lambda_i}(n),g_{\mu_i}(n), g_{\nu}(n),\break g_{\pi}(n)\}$ collects all the subgradients of \eqref{eq.probdualfunc} with respect to (w.r.t.) the Lagrange multipliers. Specifically, we have
\begin{subequations}\label{eq.subgradient}
\begin{align}
g_{\lambda_i}(n)&=\frac{a_i^3 R_i^3k_iB_i^3}{(1-\phi)^2 T^2}+\frac{\sigma^2}{g_i}(e^{\frac{(1-a_i)R_i}{T_{\text{off},i}w_i}}-1)T_{\text{off},i}
- P_{b,i} \phi T \theta_i H_i,  \\
g_{\mu_i}(n)&=T_{\text{off},i}+\frac{(1-a_i)R_iB_i}{f_{s,i}}- (1-\phi)T,  \\
g_{\nu}(n)&= \sum_i f_{s,i}- f_s^{\max},\\
g_{\pi}(n)&= \sum_i P_{b,i}- P_b^{\max}.
\end{align}
\end{subequations}

As the non-constrained problem \eqref{eq.probdualfunc} does not include any constraint or term that couples two or more optimization variables $\{{\boldsymbol f_s},{\boldsymbol P_b},{\boldsymbol T_{\text{off}}}\}$, we decompose problem \eqref{eq.probdualfunc} into the following $(2I+1)$ subproblems of different optimization variables to obtain the primal $\{{\boldsymbol f_s},{\boldsymbol P_b},{\boldsymbol T_{\text{off}}}\}$ in \eqref{eq.subgradient}:
\begin{align}\label{eq.t}
\{T_{\text{off},i}\}_{i \in {\cal I}}
&\in \arg \min_{T_{\text{off},i}}\Big[\lambda_i \frac{\sigma^2}{g_i}(e^{\frac{(1-a_i)R_i}{T_{\text{off},i}w_i}}-1)T_{\text{off},i}+\mu_i T_{\text{off},i}\Big].
\end{align}

\begin{align}\label{eq.alpha}
\{P_{b,i}\}_{i \in {\cal I}}
&\in \arg \min_{P_{b,i}}  \Big[(\phi T-\lambda_i \phi T \theta_i  H_i +\pi ) P_{b,i}\Big].
\end{align}

\begin{align}\label{eq.fs}
\{\boldsymbol f_s\}
&\in \arg \min_{f_{s,i}} \Big[\sum_i [(1-a_i)R_i\delta B_i f_{s,i}^2
+\mu_i\frac{(1-a_i)R_iB_i}{f_{s,i}}+\nu f_{s,i}]+ E_{\text{cool}}(f_{s,i})\Big].
\end{align}

When a constant stepsize $\eta$ is used, the subgradient iterations in \eqref{eq.multiplier} converges to a neighborhood of the optimal $\boldsymbol{\omega}^*$ for the dual problem from any initial point $\boldsymbol{\omega}(0)$. The size of the neighborhood is proportional to the stepsize $\eta$. If we use a set of non-summable diminishing stepsizes satisfying $\lim_{n \rightarrow \infty} \eta(n) =0$ and $\sum_{n=0}^{\infty} \eta(n) = \infty$, the iterations in \eqref{eq.multiplier} converge to the exact $\boldsymbol{\omega}^*$ as $j \rightarrow \infty$ \cite{convex}. Since \eqref{eq.prob1} is convex and satisfies the Slater's condition, the duality gap between \eqref{eq.prob1} and \eqref{eq.probdual} is zero, and the convergence to $\boldsymbol{\omega}^*$ leads to the optimal solution $\{{\boldsymbol f_s},{\boldsymbol P_b},{\boldsymbol T_{\text{off}}}\}$ to the primal problem \eqref{eq.prob1}. In what follows, given $\boldsymbol{\omega}$, we provide solutions for the offloading management subproblem \eqref{eq.t}, the WPT power control subproblem \eqref{eq.alpha} and the CPU frequency allocation subproblem \eqref{eq.fs}, respectively.

%
\subsubsection{Offloading management}
Since problem \eqref{eq.t} is convex, we can apply the Karush-Kuhn-Tucker (KKT) conditions \cite{convex} to obtain the optimal solution $T_{\text{off},i}^*$ to \eqref{eq.t} in a semi-closed form, which is stated in the following lemma.

\begin{lemma}\label{lemma3}
For any given $\boldsymbol{\omega}$, the optimal solution $T_{\text{off},i}^*$ to \eqref{eq.t} can be expressed as:
\begin{equation}\label{eq.tclose}
T_{\text{off},i}^*=\left\{
\begin{array}{ll}
    0 ,&\text{if}~\lambda_i = 0,  \\
     \frac{(1-a_i)R_i}{w_i\big[W_0\big(\frac{g_i\mu_i}{\sigma^2 e \lambda_i}-\frac{1}{e}\big)+1\big]}, &\text{if}~\lambda_i > 0,\\
\end{array}
\right.
\end{equation}
where $W_0(\cdot)$ is the principal branch of the Lambert $W$ function satisfying $W_0(x)e^{W_0(x)}=x$\cite{Corless1996On}, and $e$ is Euler's number.
\end{lemma}

\begin{IEEEproof}
For $\lambda_i=0$, the objective function of \eqref{eq.t} becomes $\mu_i T_{\text{off},i}$. It is clear that, given $\boldsymbol{\omega}$, $T_{\text{off},i}=0$ is optimal for \eqref{eq.t}.

For $\lambda_i>0$, the optimal $T_{\text{off},i}^*$ is achieved when the first-order derivative of \eqref{eq.t} is zero.
Recall \eqref{eq.perspective}, i.e., $P(\frac{(1-a_i)R_i}{T_{\text{off},i}}):=e^\frac{(1-a_i)R_i}{T_{\text{off},i}w_i}-1$. Then solving problem \eqref{eq.t} is equivalent to solving
\begin{align}\label{eq.y}
P(\frac{(1-a_i)R_i}{T_{\text{off},i}^*})-\frac{(1-a_i)R_i}{T_{\text{off},i}^*}P'(\frac{(1-a_i)R_i}{T_{\text{off},i}^*})=-\frac{g_i\mu_i}{\sigma^2\lambda_i},
\end{align}
where $P'(\cdot)$ is the derivative of $P(\cdot)$, and the left-hand side of \eqref{eq.y} is the first-order derivative of $[P(\frac{(1-a_i)R_i}{T_{\text{off},i}^*})T_{\text{off},i}^*]$ w.r.t. $T_{\text{off},i}^*$.
For the function $y=P(x)-xP'(x)$ of $x>0$, its inverse function is given by \cite{Corless1996On}
\begin{align}\label{eq.wfunc}
x=w_i\big[W_0(-\frac{y}{e}-\frac{1}{e})+1\big].
\end{align}
Let $r_i:=\frac{(1-a_i)R_i}{T_{\text{off},i}}$ denote the transmit rate of user $i$ for offloading its tasks. Based on \eqref{eq.y}, we have $y=P(r_i^*)-r_i^*P'(r_i^*)=-\frac{g_i\mu_i}{\sigma^2\lambda_i}$. It then follows that
\begin{align}\label{eq.r}
r_i^*=\frac{(1-a_i)R_i}{T_{\text{off},i}^*}=w_i\big[W_0(\frac{g_i\mu_i}{\sigma^2 e \lambda_i}-\frac{1}{e})+1\big],
\end{align}
which is immediately followed by \eqref{eq.tclose}.
\end{IEEEproof}

\subsubsection{WPT power control}
Problem \eqref{eq.alpha} is a linear program. The optimal $P_{b,i}^*$ to \eqref{eq.alpha} can also be derived in a semi-closed form, as stated in the following lemma.
\begin{lemma}\label{lemma4}
Given $\boldsymbol{\omega}$, the optimal solution $P_{b,i}^*$ to \eqref{eq.alpha} is given by
\begin{align}\label{eq.pclose}
P_{b,i}^*=\left\{
\begin{array}{ll}
     0 ,&\text{if}~(\phi T-\lambda_i \phi T \theta_i  H_i +\pi )> 0 ,\\
     P_b^{\max}, &\text{if}~(\phi T-\lambda_i \phi T \theta_i  H_i +\pi )< 0,\\
   \tilde{P_{b,i}},&\text{if}~(\phi T-\lambda_i \phi T \theta_i  H_i +\pi )=0,
\end{array}
\right.
\end{align}
where $\tilde{P_{b,i}}= [\frac{a_i^3 R_i^3k_iB_i^3}{(1-\phi)^2 T^2}+\frac{\sigma^2}{g_i}(e^{\frac{(1-a_i)R_i}{T_{\text{off},i}^*w_i}}-1)T_{\text{off},i}^*]/(\phi T \theta_i H_i)$.
\end{lemma}

\begin{IEEEproof}
It is easy to determine $P_{b,i}$ when $(\phi T-\lambda_i \phi T \theta_i  H_i +\pi )\neq 0$.
For $(\phi T-\lambda_i \phi T \theta_i  H_i +\pi )=0$, it is easy to find that $\lambda_i>0$.
Given its convexity, the complementary slackness condition holds for problem \eqref{eq.prob1} \cite{convex}, i.e., $\forall i \in {\cal I}$,
\begin{align}\label{comslack1}
\lambda_i^* [\frac{a_i^3 R_i^3k_iB_i^3}{(1-\phi)^2 T^2}+T_{\text{off},i}^*(e^{\frac{(1-a_i)R_i}{T_{\text{off},i}^*w_i}}-1)
- P_{b,i}^* \phi T \theta_i H_i]=0.
\end{align}
For $\lambda_i>0$, we have
\begin{align}\label{energy_causility_meet}
[\frac{a_i^3 R_i^3k_iB_i^3}{(1-\phi)^2 T^2}+T_{\text{off},i}^*(e^{\frac{(1-a_i)R_i}{T_{\text{off},i}^*w_i}}-1)
- P_{b,i}^* \phi T \theta_i H_i]=0.
\end{align}
Lemma~\ref{lemma4} readily follows.
\end{IEEEproof}

%

\subsubsection{CPU frequency resource allocation}
Problem \eqref{eq.fs} is also a convex function of ${\boldsymbol f_s}$. According to the definition of the cooling energy consumption in \eqref{cooling1} and \eqref{cooling2}, there is no closed-form solution for ${\boldsymbol f_s}$. We solve problem \eqref{eq.fs} by using off-the-shelf convex solvers, such as interior-point method.

After the solution ${\boldsymbol \omega}$ to the dual problem \eqref{eq.probdual} converges to the optimal one ${\boldsymbol \omega^{\text{opt}}}=\{\lambda_i^{\text{opt}},\mu_i^{\text{opt}},\break \nu^{\text{opt}},\pi^{\text{opt}},\forall i \in {\cal I}\}$ by the subgradient method, we can obtain the optimal solution $\{{\boldsymbol f_{s}}^{\text{opt}},{\boldsymbol P_b}^{\text{opt}}, \break{\boldsymbol T_{\text{off}}}^{\text{opt}}\}$ to \eqref{eq.prob1} based on the following proposition.
\begin{proposition}\label{proposition1}
Given the task ratio vector for local computing ${\boldsymbol a}$, the optimal solution $\{{\boldsymbol f_{s}}^{\text{opt}},{\boldsymbol P_b}^{\text{opt}}, \break{\boldsymbol T_{\text{off}}}^{\text{opt}}\}$ to problem \eqref{eq.prob1} is obtained when $(\phi T-\lambda_i^{\text{opt}} \phi T \theta_i  H_i +\pi^{\text{opt}} )=0$, as given by
\begin{align}
T_{\text{off},i}^{\text{opt}}&=\frac{(1-a_i)R_i}{w_i\big[W_0\big(\frac{g_i\mu_i^{\text{opt}}}{\sigma^2 e \lambda_i^{\text{opt}}}-\frac{1}{e}\big)+1\big]},
\label{eq.optimalt}\\
P_{b,i}^{\text{opt}} &= [\frac{a_i^3 R_i^3k_iB_i^3}{(1-\phi)^2 T^2}+\frac{\sigma^2}{g_i}(e^{\frac{(1-a_i)R_i}{T_{\text{off},i}^{\text{opt}}w_i}}-1)T_{\text{off},i}^{\text{opt}}]/(\phi T \theta_i H_i),
\label{eq.optimalalpha}\\
f_{s,i}^{\text{opt}} &= \frac{(1-a_i)R_iB_i}{ (1-\phi)T-T_{\text{off},i}^{\text{opt}}}, \quad \forall i \in {\cal I}.\label{eq.optimalfs}
\end{align}
\end{proposition}

\begin{IEEEproof}
See Appendix B.
\end{IEEEproof}

\subsubsection{Local/edge computing load management}
Given the optimal solution $\{{\boldsymbol f_{s}}^{\text{opt}},{\boldsymbol P_b}^{\text{opt}}, {\boldsymbol T_{\text{off}}}^{\text{opt}}\}$ to problem \eqref{eq.prob1}, as specified in Proposition~1, we proceed to calculate the optimal ratio of a task for local computing $\boldsymbol a$.
Given $\{{\boldsymbol f_s},{\boldsymbol P_b},{\boldsymbol T_{\text{off}}}\}$, problem \eqref{eq.prob} can be reduced to a convex form:
\begin{align} \label{eq.optimala}
& \mathop{\text{min}}\limits_{\boldsymbol a} \;\sum_i E_{\text{comp},i}(a_i)+E_{\text{cool}}(a_i) \\
& \text{suject to:}  \quad \eqref{eq.probb1}~\text{and}~\eqref{eq.probd1}.\notag
\end{align}
The Lagrangian function of \eqref{eq.optimala} is given by:
 \begin{align} \label{eq.optimala_lag}
&{\cal L}(\boldsymbol a)=\sum_i \Big\{(1-a_i)R_i\delta B_i f_{s,i}^2 + E_{\text{cool}}(a_i)\notag\\
&+\lambda_i [\frac{a_i^3 R_i^3k_iB_i^3}{(1-\phi)^2 T^2}+\frac{\sigma^2}{g_i}T_{\text{off},i}(e^{\frac{(1-a_i)R_i}{T_{\text{off},i}w_i}}-1)-P_{b,i} \phi T \theta_i H_i]\notag\\
& +  \mu_i[T_{\text{off},i}+\frac{(1-a_i)R_iB_i}{f_{s,i}}- (1-\phi)T]\Big\}.
\end{align}
Due to the convexity of problem \eqref{eq.optimala}, we can solve it equivalently by solving $\min_{\boldsymbol a} {\cal L}(\boldsymbol a)$. With the optimal $\{{\boldsymbol f_{s}}^{\text{opt}},{\boldsymbol P_b}^{\text{opt}}, {\boldsymbol T_{\text{off}}}^{\text{opt}},{\boldsymbol \omega^{\text{opt}}}\}$  derived by solving problems \eqref{eq.prob1} and \eqref{eq.probdual}, ${\cal L}(\boldsymbol a)$ is convex in $\boldsymbol a$, and the optimal $\boldsymbol a^{\text{opt}}$ can be solved by standard convex solvers, such as interior-point method.

Given the optimal $\boldsymbol a^{\text{opt}}$, we solve problem \eqref{eq.prob1} and then $\min_{\boldsymbol a} {\cal L}(\boldsymbol a)$ iteratively, until the convergence to a locally optimal solution to the original problem \eqref{eq.prob}. The proposed algorithm is summarized in Algorithm 1.

\begin{algorithm}[t]
\caption{Proposed Joint Design for WPT-MEC}
\begin{algorithmic}[1]
\For {$j=1,2,\ldots$}
\State Given $\boldsymbol a(j)$, solve problem \eqref{eq.prob1}, and obtain the optimal $\{{\boldsymbol f_{s}}(j),{\boldsymbol P_b(j)}, {\boldsymbol T_{\text{off}}}(j)\}$ in a semi-closed form, as in \eqref{eq.optimalt}-\eqref{eq.optimalfs}.

\State Given $\{{\boldsymbol f_{s}}(j), {\boldsymbol P_b}(j), {\boldsymbol T_{\text{off}}}(j)\}$, solve problem \eqref{eq.optimala}, and obtain the optimal ratio of a task for local computing ${\boldsymbol a}(j+1)$.
\EndFor
\State Repeat the above steps until $\{{\boldsymbol f_{s}},{\boldsymbol P_b}, {\boldsymbol T_{\text{off}}}, {\boldsymbol a}\}$ converge within a predefined accuracy.

\end{algorithmic}
\end{algorithm}


\section{Properties and Insights}
Proposition 1 reveals that, given the ratio of the locally computing part of every task, the proposed joint resource allocation and load management design of the cooling-aware WPT-MEC system has the following interesting design insights.

\begin{enumerate}
%
\item For each user $i$, if the latency constraint \eqref{eq.probd1} is more stringent, the tasks are offloaded to the server within a shorter offloading time ${T_{\text{off},i}^{\text{opt}}}$. This can be verified by the complementary slackness condition of problem \eqref{eq.prob1}, that is,
\begin{align}\label{comslack2}
\mu_i^{\text{opt}}[T_{\text{off},i}^{\text{opt}}+\frac{(1-a_i)R_iB_i}{f_{s,i}^{\text{opt}}}- (1-\phi)T]=0, ~~ \forall i \in {\cal I},
\end{align}
which dictates that a more stringent latency constraint (e.g. with a smaller $T$ or a large $\phi$) admits a larger $\mu_i^{\text{opt}}$, consequently leading to a shorter offloading time ${T_{\text{off},i}^{\text{opt}}}$ according to \eqref{eq.optimalt} and a higher offloading rate $r_i^{\text{opt}}$.

\item The offloading time ${T_{\text{off},i}^{\text{opt}}}$ would decrease when i) the channel condition from the AP to user $i$ becomes better (i.e., $H_i$ becomes larger), ii) the energy conversion efficiency $\theta_i\in (0,1)$ of user $i$ becomes larger.
As revealed in \eqref{eq.optimalt}, a smaller $\lambda_i^{\text{opt}}$ admits a shorter offloading time ${T_{\text{off},i}^{\text{opt}}}$ and a lower transmit rate $r_i^{\text{opt}}$. Based on Proposition 1, the optimal solution is obtained when $(\phi T-\lambda_i^{\text{opt}} \phi T \theta_i  H_i +\pi^{\text{opt}} )=0$. Therefore, a large $H_i$ or $\theta_i$ results in a smaller $\lambda_i^{\text{opt}}$,  and eventually a shorter offloading time ${T_{\text{off},i}^{\text{opt}}}$. 


\item The offloading time ${T_{\text{off},i}^{\text{opt}}}$ would also decrease when the offloading efficiency is improved, e.g., the channel condition from user $i$ to the AP improves or the bandwidth of user $i$ becomes wider. The transmit power $P_{b,i}^{\text{opt}}$ for WPT would decrease when the WPT efficiency is improved, e.g., the channel condition from the AP to user $i$ improves or the energy conversion efficiency $\theta_i\in (0,1)$ of user $i$ increases.


\end{enumerate}

\section{Numerical results}

In this section, numerical tests are presented to evaluate the proposed design of joint resource allocation and load management in WPT-MEC systems. The default setting of the system parameters in our experiments is listed in Table~\ref{table.parameter_setting}. 
\subsection{Experiment Setup}
\renewcommand\arraystretch{1.2}
\begin{table*}[h]
	\caption{Parameters for performance analysis}
	\begin{center}
		\begin{tabular}{p{7cm}|p{2cm}}
		
			\hline
			Parameters & Values\\
		        \hline
			\hline
			Time splitting ratio for WPT $\phi$ & 0.4\\
			\hline
			Maximum CPU frequency of the server $f_s^{\max}$ & 2 GHz\\
			\hline 	
			Maximum CPU frequency of user $i$, $f_{u,i}^{\max}$ & 1 GHz\\
			\hline
			Required CPU cycles for per nat computing $B_i$ & $10^3$ \\
			\hline
			Effective switched capacitance coefficients $k_i$ and $\delta$& $10^{-26}$ \\
			\hline
			Noise power at the receiver of the AP $\sigma^{2}$ & $10^{-9}$ \\
			\hline
			Energy conversion efficiency $\theta_i$ & $0.3$ \\
			\hline
			Effective channel gain from the AP to user $i$, $H_i$ & $10^{-3}$ \\
			\hline
			Maximum transmit power of the AP $P_{b}^{\max}$ & 20 W\\
			\hline
 			System bandwidth $W$ & 5 MHz\\
			\hline
			Task size of user $i$ $R_i$ & 1.5 Knats\\
			\hline
		\end{tabular}
		\label{table.parameter_setting}
	\end{center}
\end{table*}


To benchmark the performance of the proposed algorithm, we evaluate three other baseline schemes, namely, local computing, full offloading, and half offloading.
\begin{enumerate}
	\item Local computing: Each user $i$ executes all tasks locally where the local computing ratio $a_i$ is set to 1. This benchmark solves problem \eqref{eq.prob1} given $a_i=1, \forall i\in \cal I$.
	\item Full offloading: Each user $i$ offloads all tasks for edge computing by setting $a_i=0, \forall i\in \cal I$.
	\item Half offloading: Each user $i$ offloads half of its tasks for edge computing by setting $a_i=0.5, \forall i\in \cal I$.
\end{enumerate}

\subsection{Effect of Task Size}
\begin{figure}
\centering
\includegraphics[width=0.6\textwidth]{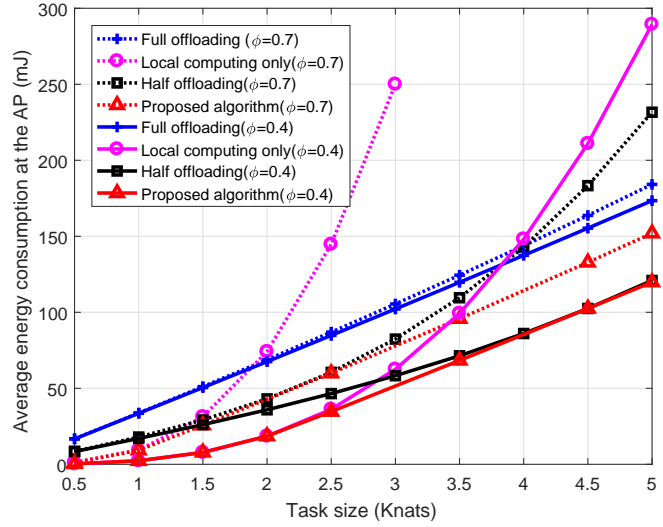}
\caption{The average energy consumption of the AP versus the task size $R_i$, where $T=0.2$ s, $I=5$ and $W=5$ MHz.}
\label{opt_vs_R}
\end{figure}
Fig. \ref{opt_vs_R} plots the average energy consumption of the AP under different schemes, as the task size $R_i$ grows. As expected, the energy consumptions of all schemes increase with the growth of the task size. It is observed that the proposed algorithm outperforms the other three schemes with fixed task ratios for local/edge computing by always incurring the lowest energy consumption. 
We also observe that the local computing scheme achieves the same performance as the proposed one when the task size is small. Yet, as the task load becomes heavier at the users, the energy consumption of the local computing scheme increases rapidly. By offloading an optimal ratio of tasks for edge computing, the energy consumption of the AP can be substantially reduced by the proposed algorithm.

The average energy consumption also increases as the time splitting ratio for WPT $\phi$ becomes larger, as shown in Fig. \ref{opt_vs_R}. This is because when $\phi$ grows, less time is assigned for offloading and local/edge computing (i.e., the latency constraint \eqref{eq.probd1} becomes more stringent), which results in higher CPU frequencies and offloading power, and consequently higher energy consumption. When $\phi=0.7$ and $R_i>3$ Knats, the ``local computing only'' scheme becomes infeasible since the limited energy supply to the users cannot support the increasing task load. In contrast, if the latency constraint is not tight (i.e., when $\phi=0.4$ s), the users can scavenge sufficient RF energy from the AP with a higher WPT transmit power. Nevertheless, $\phi$ cannot be too small, as the WPT transmit power is limited by the AP's maximum transmit power.   

\begin{figure}
\centering
\includegraphics[width=0.6\textwidth]{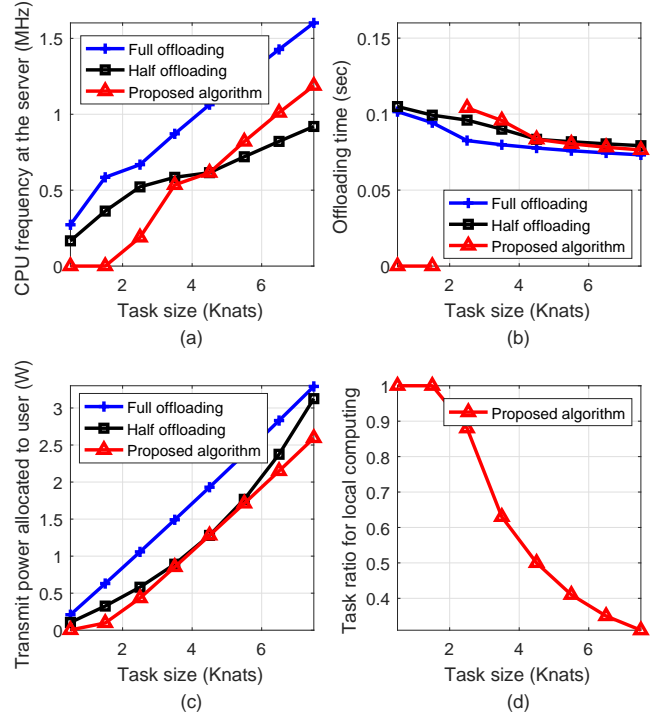}
\caption{The CPU frequency at the MEC server $f_{s,i}$, the offloading time ${T_{\text{off},i}}$, the transmit power for WPT $P_{b,i}$, and the task ratio for local computing $a_i$ versus the task size $R_i$, where $T=0.2$ s, $I=5$, $W=5$ MHz and $\phi=0.4$.}
\label{3_vs_R}
\end{figure}

Fig.~\ref{3_vs_R} depicts the CPU frequency at the MEC server $f_{s,i}$, the offloading time ${T_{\text{off},i}}$, the transmit power for WPT $P_{b,i}$, and the task ratio for local computing $a_i$ of user $i$, as the task size $R_i$ grows.
As shown in Fig.~\ref{3_vs_R}(a), as the task load becomes heavier, the CPU frequencies of all schemes increase to improve the task execution capability at the server, so that all the incoming tasks can be executed before their deadlines.
In addition, since the MEC server employing the full offloading scheme always performs more task executions than the other two schemes, the full offloading scheme has to apply the highest CPU frequency.

In Fig.~\ref{3_vs_R}(b), we see that the offloading times of all schemes decrease as $R_i$ grows, among which the offloading time of the full offloading scheme is the shortest. This is because the increased amount of offloading data (i.e., $(1-a_i)R_i$) makes the latency constraint \eqref{eq.probd1} more stringent, leading to a shorter offloading time and a higher transmit rate. This corroborates the property of the proposed algorithm stated in Section IV. It is shown in Fig.~\ref{3_vs_R}(c) that the transmit powers for WPT increase as $R_i$ grows. This is intuitive: As more offloading and local computing energy is consumed for heavier task load, a higher transmit power should be employed to provide a higher energy supply for users.

It can be seen from Fig.~\ref{3_vs_R}(d) that at small $R_i$ values (e.g., when $R_i$ is smaller than 1.5 Knats), the ratio of a task for local computing remains 1; in other words, the proposed algorithm only performs local computing. This is also the reason for which the CPU frequency and the offloading time of the proposed algorithm remain 0 at small $R_i$ values in Figs.~\ref{3_vs_R}(a) and~\ref{3_vs_R}(b).
As the task size grows, the proposed algorithm offloads more data for edge computing, since the energy consumption per nat for offloading becomes lower than that for local computing.

\begin{figure}
\centering
\includegraphics[width=0.6\textwidth]{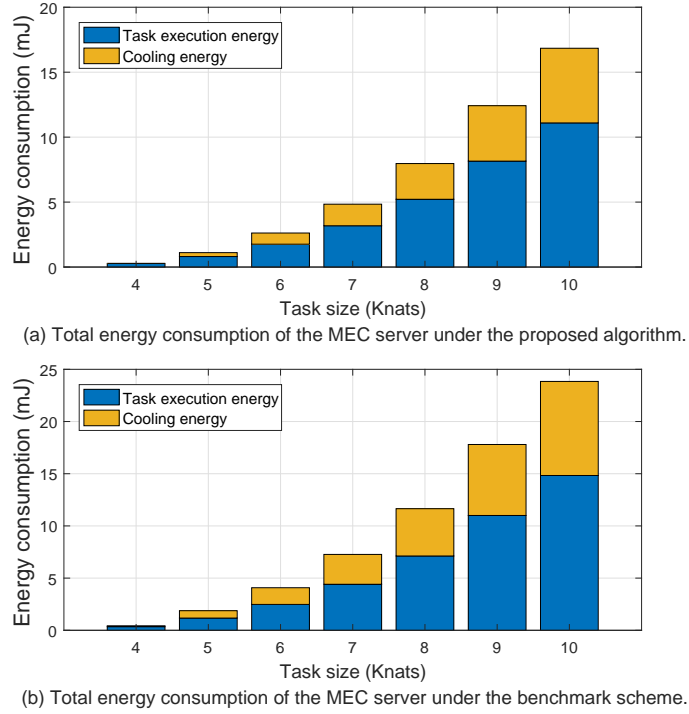}
\caption{The energy consumption for task execution and cooling at the server versus the task size $R_i$, where $T=0.2$ s, $I=5$, $W=5$ MHz and $\phi=0.4$.}
\label{IT_Cooling_vs_B}
\end{figure}

To demonstrate the merit of incorporating cooling energy models
in our proposed joint design of the WPT-MEC system, we
compare the proposed algorithm with a benchmark
scheme, where the effects of cooling energy consumption are overlooked in the energy minimization problem. The total energy consumption of the MEC server, including the task
execution energy and the cooling energy, is plotted in Fig. \ref{IT_Cooling_vs_B}, as
the task size $R_i$ grows. Obviously, as the proposed algorithm
jointly manages edge computing and cooling activities for
energy efficiency, both the task execution energy and the
cooling energy are lower under the proposed algorithm than
those under the benchmark scheme. As expected, the
trend of the computing or cooling energy consumption
is consistent with that of $f_{s,i}$ (which increases as the task
size grows; see Fig. \ref{3_vs_R}(a)), since the computing or the cooling
energy is an increasing function of $f_{s,i}$. We also see
that the proportion of the cooling energy in the total energy
consumption at the server grows up, as $R_i$  increases. This is
due to the optimization of the cooling energy \eqref{cooling1} applied in
our proposed algorithm. The OA cooling, which causes less
energy consumption than the CW cooling, is first adopted until
the total computing power exceeds the threshold $P_\text{TH}$. With the
growth of the total computing power, the cooling energy consumption
is gradually and increasingly dominated by the CW cooling, leading
to an increasing proportion in the total energy consumption.

\subsection{Effect of User Number and Slot Duration}
\begin{figure}
\centering
\includegraphics[width=0.6\textwidth]{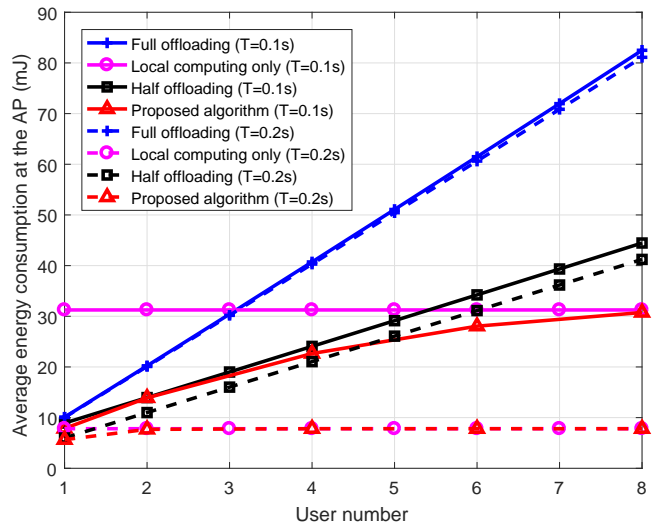}
\caption{The average energy consumption of the AP versus the user number $I$, where $W=5$ MHz, $\phi=0.4$ and $R_i=1.5$ Knats.}
\label{opt_vs_num}
\end{figure}

Fig. \ref{opt_vs_num} plots the average energy consumption of the AP, as the user number grows. 
It can be seen that the proposed algorithm always outperforms the other three schemes, and the gains of the proposed approach (compared to the full and half offloading schemes) enlarge as the user number grows.
In particular, when $T=0.2$~s and the user number is 8, the proposed algorithm can save up to 90.4\% and 81.0\% of the energy, as compared to the full offloading and half offloading schemes, respectively.
It is also observed that, with a larger time slot duration $T$, all schemes can reduce the energy consumption. This is because a longer time slot allows more time for offloading and computing, so that lower CPU frequency and offloading power can be employed, leading to a lower energy consumption.

We also see that the performance of the proposed algorithm approaches that of the local computing only scheme, with growing number of users. This is because the optimal local computing ratio $a_i$ converges to 1 by our proposed algorithm. The growth of the user number leads to smaller uplink bandwidth assigned to each user, so that the computation offloading takes longer time, and the energy consumption for edge computing increases due to less time allocated. As the bandwidth per user becomes smaller, the users choose not to offload and compute all tasks locally to save energy. The average energy consumption of the local computing only scheme remains unchanged as the local computing energy of each user is unaffected by the user number (or uplink bandwidth).

\begin{figure}
\centering
\includegraphics[width=0.6\textwidth]{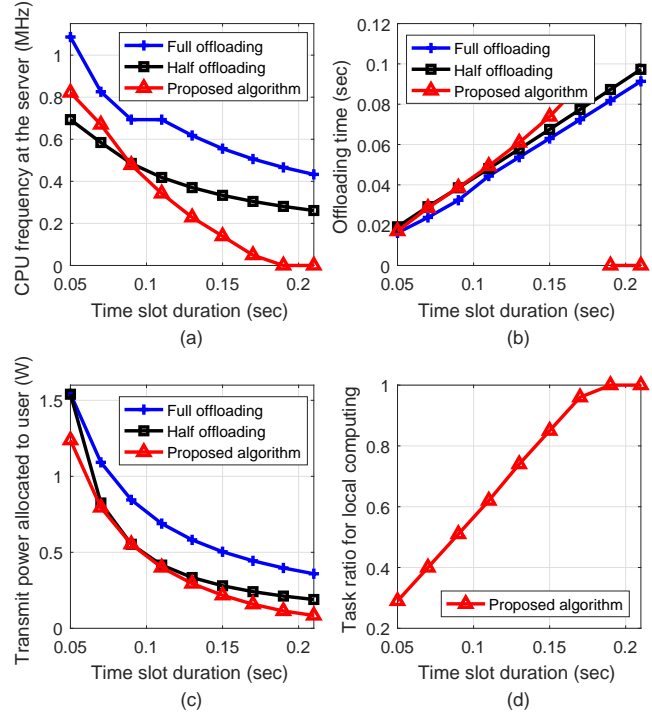}
\caption{The CPU frequency at the MEC server $f_{s,i}$, the offloading time ${T_{\text{off},i}}$, the transmit power for WPT $P_{b,i}$, and the task ratio for local computing $a_i$ versus the time slot duration $T$, where $I=3$, $W=5$ MHz, $\phi=0.4$ and $R_i=1.5$ Knats.}
\label{3_vs_T}
\end{figure}

Fig.~\ref{3_vs_T} shows the CPU frequency at the MEC server $f_{s,i}$, the offloading time ${T_{\text{off},i}}$, the transmit power for WPT $P_{b,i}$, and the task ratio for local computing $a_i$, as the time slot duration $T$ becomes large.
It is shown in Fig.~\ref{3_vs_T}(a) that the CPU frequencies of all schemes decline with the growth of the time slot duration $T$.
This result is intuitive: The growth of the time slot duration allows a looser latency requirement, which in turn, leads to a lower CPU frequency to save energy.
Moreover, the offloading times of different schemes in Fig.~\ref{3_vs_T}(b) increase as $T$ grows. This is because, as the latency constraint becomes looser, more time is available to be assigned for computation offloading with a lower transmit rate.
In Fig.~\ref{3_vs_T}(c), the transmit powers for WPT decline as $T$ grows. This is because the energy consumed by offloading and local computing decreases with a large $T$.
It is also obvious that when $T\geq 0.19$ s, the proposed algorithm preferably performs the local computing only scheme. In other words, the users would offload more data when the latency constraint becomes more stringent.


\subsection{Impact of Bandwidth}
\begin{figure}
\centering
\includegraphics[width=0.6\textwidth]{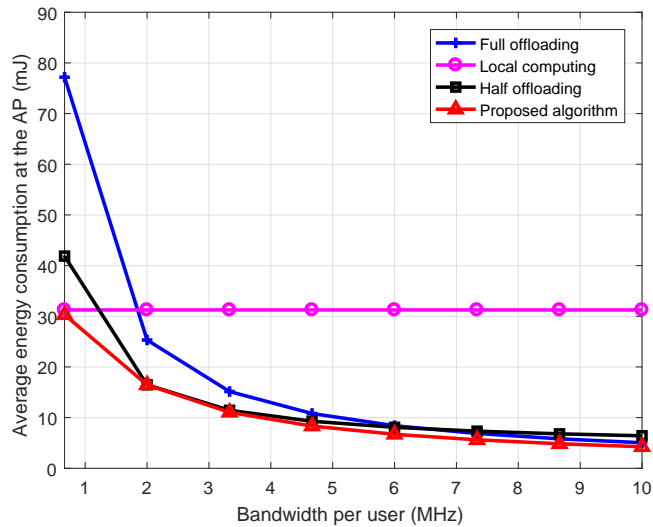}
\caption{The average energy consumption of the AP versus the bandwidth per user $w_i$, where $T=0.1$ s, $I=3$, $\phi=0.4$ and $R_i=1.5$ Knats.}
\label{opt_vs_B}
\end{figure}

Fig. \ref{opt_vs_B} plots the average energy consumption of the AP versus the bandwidth per user $w_i$. 
As expected, the energy consumptions of the schemes with offloading operations decline with the growth
of $w_i$. Relying on \eqref{eq.optimalt}, a larger bandwidth can reduce the offloading time and save more time for edge computing, thus reducing energy consumption. The energy for local computing is independent of the bandwidth; the same conclusion has been claimed in Fig. \ref{opt_vs_num}.

\begin{figure}
\centering
\includegraphics[width=0.6\textwidth]{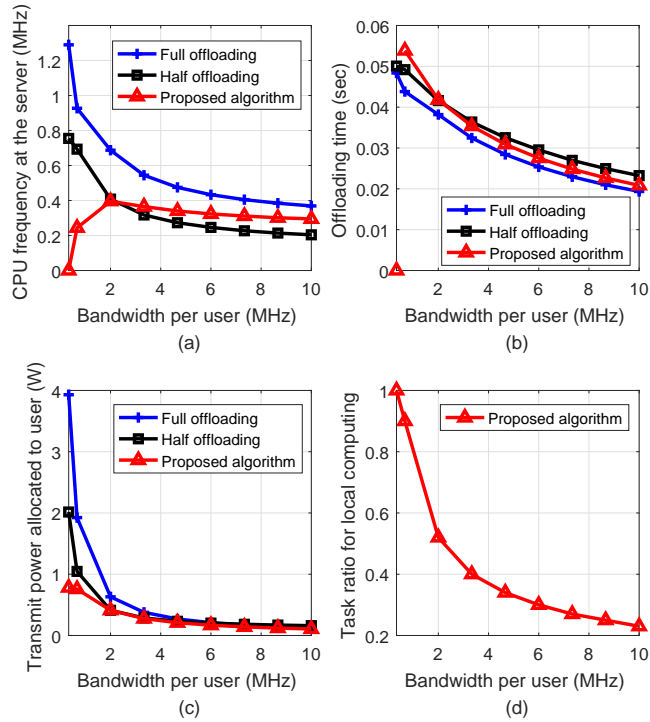}
\caption{The CPU frequency at the MEC server $f_{s,i}$, the offloading time ${T_{\text{off},i}}$, the transmit power for WPT $P_{b,i}$, and the task ratio for local computing $a_i$ versus the bandwidth per user $w_i$, where $T=0.1$ s, $I=3$, $\phi=0.4$ and $R_i=1.5$ Knats.}
\label{3_vs_B}
\end{figure}
Fig.~\ref{3_vs_B} depicts the CPU frequency at the MEC server $f_{s,i}$, the offloading time ${T_{\text{off},i}}$, the transmit power for WPT $P_{b,i}$, and the task ratio for local computing $a_i$, as the bandwidth per user $w_i$ increases.
In Fig.~\ref{3_vs_B}(a), we see that the CPU frequency at the server of the full or half offloading scheme decreases as the bandwidth assigned to each user grows.
As a larger bandwidth can save more time for edge computing, given task amount, the CPU frequency at the edge is lowered. This is also corroborated by Fig.~\ref{3_vs_B}(b).
In Fig.~\ref{3_vs_B}(c), the transmit powers for WPT decline because the energy consumption for offloading is reduced with a larger uplink bandwidth, and that for local computing stays unchanged.
From Fig.~\ref{3_vs_B}(d), we observe that in the proposed algorithm, the offloading task amount rises when a user is assigned a larger bandwidth. As a result, $f_{s,i}$ under the proposed algorithm first increases and then slightly decreases, as shown in Fig.~\ref{3_vs_B}(a).

\begin{table*}[h]
\renewcommand{\arraystretch}{1.5}
\centering
	\caption{Offloading time and transmit power for WPT of the proposed joint design under different $H_i$ and $\theta_i$ }
	\begin{center}
		\begin{tabular}{ c |c|c|c|c|c|c|c|c|c}
		
			\hline
			\multicolumn{2}{c|}{Channel gain $H_i (\times10^{-3})$} & 1&2&3&4&5&6&7&8\\\hline
			
			\multirow{2}*{Offloading time ${T_{\text{off},i}}$ (ms)} & $\theta_i=0.3$ &   95.7&90.6&87.2&84.6&82.5&80.7&79.2&77.8\\\cline{2-10}
			
			~&$\theta_i=0.6$ &   90.6&84.6&80.7&77.8&75.5&73.5&71.8&70.4\\\cline{1-10}
				
			\multirow{2}*{Transmit power for WPT $P_{b,i}$ (mW)}&  $\theta_i=0.3$ &   853.4&426.8&284.6&213.4&170.8&142.3&122.0&106.7\\\cline{2-10}
			
             ~ & $\theta_i=0.6$&  426.8&213.4&142.3&106.7&85.4&71.2&61.0&53.4\\\cline{2-10}
			\hline
		\end{tabular}
		\label{table.t-p-H}
	\end{center}
\end{table*}

Finally, we corroborate the properties obtained in Section IV by the results in Table~\ref{table.t-p-H}, where the offloading time ${T_{\text{off},i}}$ and the transmit power $P_{b,i}$ for WPT decrease when i) the channel condition from the AP to the user improves (i.e., $H_i$ becomes larger), and ii) the energy conversion efficiency $\theta_i\in (0,1)$ of user $i$ increases. 




\section{Conclusion}
A joint design of resource allocation and load management for cooling-aware WPT-MEC systems was developed in this paper. We minimized the total energy consumption of the AP, by jointly optimizing the transmit power for WPT, the local/edge computing load, the offloading time, and the CPU frequencies at the AP and the users. By rigorously orchestrating the alternating optimization technique, Lagrange duality and subgradient descent methods, we decomposed the original optimization problem and obtained the optimal solution in a semi-closed form.
It was revealed that the users would offload more data when the latency requirements become stringent.
Extensive numerical tests corroborated that the proposed joint design can substantially save energy consumption of the AP over benchmark schemes. 

\appendix

\subsection{Proof of Lemma~\ref{lemma2}}
From \eqref{Tloc}, we can find that the longer the local task execution time $T_{\text{loc},i}$ is, the lower the local CPU frequency ${\boldsymbol f_u}$ is. As a result, less energy is to be consumed for local computing. According to \eqref{eq.probb}, the energy consumption for local computing should be as low as possible, such that the energy consumption of the AP for WPT in the objective \eqref{eq.proba} can be reduced. As $T_{\text{loc},i}$ is upper-bounded by $(1-\phi)T$ in \eqref{eq.probe}, it follows that
\begin{equation}
T_{\text{loc},i}=\frac{a_iR_i B_i}{f_{u,i}}=(1-\phi)T.
\label{f_u1}
\end{equation}
By combining \eqref{eq.probf} and \eqref{f_u1}, Lemma \ref{lemma2} is proved.

\subsection{Proof of Proposition~\ref{proposition1}}

Given the task ratio vector for local computing ${\boldsymbol a}$, the energy causality constraint \eqref{eq.probb1} should meet with equality to minimize the total energy consumption at the AP. This indicates that $\lambda_i^{\text{opt}}>0$ and $(\phi T-\lambda_i^{\text{opt}} \phi T \theta_i  H_i +\pi^{\text{opt}} )=0$. \eqref{eq.optimalt} and \eqref{eq.optimalalpha} can be derived directly from Lemmas~\ref{lemma3} and \ref{lemma4}.

Eq. \eqref{eq.optimalfs} can be proved in the same way as Lemma~\ref{lemma2}, i.e., the CPU frequency $f_{s,i}$ satisfying $T_{\text{off},i}+\frac{(1-a_i)R_iB_i}{f_{s,i}} = (1-\phi)T$ is the optimal solution to problem \eqref{eq.prob1}.

\bibliographystyle{IEEEtran}
\bibliography{references}
\end{document}